\documentclass[10pt,twocolumn,letterpaper]{article}

\usepackage[margin=0.75in,columnsep=0.25in]{geometry}
\usepackage{setspace}
\usepackage{array}
\usepackage{tcolorbox}

\usepackage{microtype}
\usepackage{amsmath,amssymb,amsfonts,amsthm}
\usepackage{booktabs} 
\usepackage{lmodern}

\usepackage{authblk}

\usepackage[numbers,sort&compress]{natbib}
\usepackage[colorlinks=true,linkcolor=blue,citecolor=blue,urlcolor=blue]{hyperref}
\usepackage{tcolorbox}

\newcounter{summarybox}
\renewcommand{\thesummarybox}{Box~\arabic{summarybox}}

\usepackage{titlesec}
\titlespacing*{\section}{0pt}{1.2ex plus 1ex minus .2ex}{0.5ex plus .2ex}
\titlespacing*{\subsection}{0pt}{1ex plus 1ex minus .2ex}{0.3ex plus .2ex}
\title{\textbf{Work, Wellbeing, and Choice:\\Empirical Lessons for AI Futures}}

\author[1]{Stephanie C.Y. Chan}
\author[1]{Adam Bales}
\author[1]{Katherine L. Hermann}
\author[1]{Iason Gabriel}

\affil[1]{Google DeepMind}

\date{} 

\begin{document}
\maketitle
\begingroup
\renewcommand{\thefootnote}{}
\footnotetext{The views and opinions expressed in this article are those of the authors and do not necessarily reflect the official policy or position of Google DeepMind or Alphabet.}
\endgroup

\begin{abstract}
Advances in AI-driven automation have raised questions about how humans might find wellbeing in a world where paid employment is less necessary or less available than before. Paid work has been variously characterized as both a contributor and an impediment to human wellbeing. What is already known about the relationship between paid work and wellbeing? What factors influence wellbeing among people who do not work---or who do not need to work? And how might these factors bear upon prospective AI-induced economic transformations? To help provide empirical grounding for these questions, we survey the psychological, sociological, and economic literature that investigates the relationship between wellbeing and work. We draw on evidence from multiple populations, including the unemployed, retirees, lottery winners, and financially dependent spouses. This comparative review draws from studies across OECD countries, China, India, and Gulf states. We identify three key factors that mediate the relationship between work status and wellbeing: (1) \textit{agency and choice}---whether the exit from work is voluntary or involuntary, as well as long-term agency; (2) the availability of \textit{alternative sources of work's latent benefits}---such as volunteering, hobbies, or state-provisioned employment; and (3) \textit{social and systemic context}---including cultural norms around work and the robustness of social safety nets. We draw on these three factors to derive specific implications for different AI automation scenarios, connecting the empirical evidence to concrete policy considerations.
\end{abstract}

\begin{table*}[!ht]
\centering
\renewcommand{\arraystretch}{1.5}
\begin{tabular}{>{\raggedright\arraybackslash}p{0.12\textwidth} p{0.22\textwidth} p{0.22\textwidth} p{0.3\textwidth}}
\toprule
\textbf{Population} & \textbf{Key characteristics} & \textbf{Potentially relevant AI scenarios} & \textbf{Key empirical takeaways} \\
\midrule

\textbf{Unemployed}\newline{\footnotesize\S\ref{sec:unemployment_costs}, \S\ref{sec:societal_systemic}} & Individuals who are jobless, available to work, and actively searching for a job. & Involuntary job displacement and work deprivation, especially in scenarios of \textit{uneven} job displacement. & Affective wellbeing often remains stable, but involuntary job loss significantly harms life satisfaction, mental health, physical health, and mortality. These effects can be long-lasting.
\vspace{4pt}
\newline
However, severity of impacts are heavily mediated by both social safety nets and cultural ``norms to work''. E.g., when the unemployed reach retirement age, they can experience increases in wellbeing, even without any change in income or routine.\\

\textbf{Lottery Winners}\newline{\footnotesize\S\ref{sec:preference_for_work}} & Experience sudden, random exogenous wealth shocks. & Sudden wealth shocks, post-abundance futures, and large-scale financial transfers. & The vast majority of lottery winners choose to keep working, even with large wins. This underscores the non-financial value of work. \\

\textbf{Retirees}\newline{\footnotesize\ref{sec:retiree_wellbeing}--\ref{sec:alternate_activities}, \ref{sec:eudaimonic_benefit}--\ref{sec:cognitive_benefit}, \ref{sec:voluntary_vs_involuntary_retirement}} & Exit the workforce, typically after a full career; tracked extensively in global longitudinal panel datasets. & Long-term post-work life, voluntary leisure, and wellbeing without employment. & Subjective wellbeing is generally high---as long as the transition is voluntary. However, this must be weighed against established risks like cognitive decline and a potential small drop in eudaimonic wellbeing (e.g., loss of purpose). Alternate activities (e.g., volunteering) help mitigate these risks and support overall wellbeing. \\

\textbf{Gulf State Citizens}\newline{\footnotesize\ref{sec:state_employment}, \ref{sec:financial_dependence}}& Historically experienced well-paid public sector employment.\footnotemark & State-provisioned or guaranteed jobs, and robust state-provided welfare systems. & State-provided employment can sustain high wellbeing, but they can create systemic vulnerabilities (e.g., high reservation wages) and a potentially risky dependency. \\

\textbf{Financially Dependent Spouses}\newline{\footnotesize\ref{sec:stay_at_home_mothers}, \ref{sec:financial_dependence}} & Individuals whose material needs are met by a partner, often trading labor force participation for domestic or caregiving roles. & Scenarios involving human dependency for material needs (e.g., relying entirely on centralized AI production or the state). & Again, wellbeing is determined more by voluntariness and alignment with personal preference, rather than work status \textit{per se}. However, financial dependence itself can be linked to reduced agency (e.g., lower ability to exit unsatisfactory relationships) and long-term vulnerability (e.g., if the provision ends). \\
\bottomrule
\end{tabular}
\caption{Overview of populations reviewed and key takeaways.}
\label{table:populations}
\end{table*}


\section{Introduction}

Recent discussions about artificial intelligence (AI) and the economy have sparked fresh interest in the relationship between work and wellbeing. According to some economic projections, advanced forms of AI could lead to an abundant automated future where many individuals are no longer obligated to work \citep{jones_i_2026}, or to a world where work is unavailable because most economic tasks are no longer performed by humans \citep{restrepo_we_2025}. Against this backdrop, two perspectives have started to emerge \citep{nyholm_meaning_2023}. At one end, there is a concern that AI could create a ``meaningfulness gap'' by taking over tasks that humans find fulfilling or enjoyable. At the other end, there is hope that advanced AI will automate tedious or meaningless tasks, which anthropologist David Graeber famously argued characterize many jobs \citep{graeber_bullshit_2018}. This could, in turn, give people time to engage in more valuable or purposeful pursuits, leading to levels of wellbeing that were previously out of reach.

These themes are also at the forefront of public concern and conversations about AI. For example, a recent U.S. poll found widespread doubt about the possibility of wellbeing without work, with 67\% of respondents stating that people need jobs in order to have purpose and dignity \citep{ramamurti_how_2025}.

However, these debates largely overlook a wealth of existing evidence on the wellbeing of populations that already live without traditional employment. We aim to bridge this gap, to provide an \textit{empirical} grounding for these discussions. This paper reviews a large body of empirical studies on non-working populations and highlights relevant findings for a world with advanced AI. We ask: What is already known about the relationship between paid work and wellbeing? What factors influence wellbeing among people who do not work---or who do not need to do so? And how might these factors bear upon prospective AI-induced economic transformations?

The degree of AI automation and the shape of future economies remains highly uncertain. Some economists and computer scientists have argued that the economic impact of advanced AI may be relatively modest or gradual \citep{acemoglu_simple_2025, narayanan_2025}. However, there are also forecasts that point to more profound structural transformation \citep{Trammell2026-yg,susskind_world_2020,jones_i_2026,restrepo_we_2025}. In this paper, we are particularly concerned with human wellbeing under scenarios of the latter kind. Relevant future scenarios depend upon the kind of wealth that AI-driven productivity gains do or do not unlock, and the way in which these gains are distributed (if they do occur). Such scenarios, which are not mutually exclusive, may include: extreme material abundance where work is available but not necessary \citep{restrepo_we_2025,korinek_preparing_2024}; sweeping financial redistribution e.g. via universal basic income (UBI) \citep{korinek_preparing_2024}; state-guaranteed or state-provisioned jobs programs \citep{tcherneva_job_2018,tymoigne_job_2013}; pervasive automation and job loss leading to widespread unemployment \citep{restrepo_we_2025,korinek_preparing_2024}; or uneven automation, where some sectors experience rapid transformation while others remain relatively untouched \citep{dellacqua_navigating_2026,acemoglu_building_2026,autor_expertise_2025}.

\footnotetext{In 2001, most Gulf State countries employed \textgreater60\% of the national labor force in the public sector (reaching 90\% in the U.A.E.) \citep{fasano_emerging_2004}.}

By reviewing the relevant psychological, sociological, and economic literature, we hope to provide an empirical basis for understanding how human wellbeing may change or evolve across these scenarios. Insights from populations that do not work, or who are not required to work, such as the unemployed, retirees, lottery winners, Gulf State citizens, and financially dependent spouses, may be especially illuminating in this context (Table \ref{table:populations}).

Overall, the involuntary loss of work tends to have a negative impact on the wellbeing of the unemployed (\S\ref{sec:unemployment_costs}). This finding is consistent with financial and non-financial benefits of work that have been documented in the literature (\S\ref{sec:benefits_of_work}). It also consistent with the finding that people often choose to work even when it is less financially necessary, e.g., in the case of lottery winners (\S\ref{sec:preference_for_work}). These results may generate pessimism about human wellbeing in futures where work is significantly automated. However, we also find populations that achieve high levels of wellbeing without working in traditional free-market jobs, such as retirees or participants in certain state-provisioned jobs; in many cases, we also find that alternate activities can provide similar benefits to those that work provides (\S\ref{sec:decoupling}). To help understand why retirees and the unemployed have such divergent outcomes, we highlight the importance of agency and choice in work status as significant determinants of wellbeing, as well as the risks of dependence (\S\ref{sec:agency_and_choice}). 
Finally, we underline the impact of societal and systemic factors, such as norms around work and the existence of social safety nets, as another set of factors explaining variation among different populations (\S\ref{sec:societal_systemic}). We close with a discussion of limitations and a summary of major takeaways (\S\ref{sec:limitations}--\ref{sec:conclusions}), which are briefly outlined in \ref{box:major_lessons}.

\refstepcounter{summarybox}\label{box:major_lessons}
\begin{tcolorbox}[float,colback=black!2!white,colframe=black!70,title=\textbf{\thesummarybox: Major Lessons for AI Economic Transitions},arc=2mm,boxrule=0.7pt,toptitle=2pt,bottomtitle=2pt]
\small
For scenarios involving significant economic transformation, the empirical evidence reviewed here points to several major lessons for wellbeing:
\begin{enumerate}\setlength{\itemsep}{2pt}\setlength{\parskip}{0pt}
    \item \textbf{Prioritize agency and choice.} Ensure short-term agency during economic transitions (because \textit{involuntary} changes in work status lead to significant loss of wellbeing). And ensure long-term agency (to avoid the risks of dependency).  (\S\ref{sec:agency_and_choice})
    
    \item \textbf{Furnish alternatives to substitute for the non-financial benefits of work}.  Existing alternatives include e.g. volunteering, civic roles, or state-provisioned job programs (though the latter must be balanced against dependency risks). (\S\ref{sec:benefits_of_work})

    \item \textbf{Account for entrenched societal norms}, such as the moral imperative to work. Large-scale shifts in work patterns can lead to cultural shifts. However, traditional work norms are likely to persist in the short-term or under \textit{uneven} displacement, harming the wellbeing of those without jobs. (\ref{sec:social_norms}--\ref{sec:others_unemployment})
    
    \item \textbf{Ensure robust safety nets} and institutional support to protect both material livelihoods and psychological wellbeing. Because the psychological and physical toll of job loss begins immediately and can leave permanent ``scars,'' interventions should be swift and proactive. (\ref{sec:safety_nets})
\end{enumerate}
\vspace{2pt}
Together, the evidence also indicates that \textbf{purely material interventions like UBI are valuable but highly insufficient.}
\noindent\textit{Further synthesis in \S\ref{sec:conclusions}:Conclusions.}
\end{tcolorbox}

\begin{table*}[htb]
\centering
\renewcommand{\arraystretch}{1.5}
\begin{tabular}{p{0.25\textwidth} p{0.33\textwidth} p{0.33\textwidth}}
\toprule
\textbf{Dimension} & \textbf{Definition} & \textbf{Examples of Measures \& Tests} \\
\midrule
\textbf{Subjective Wellbeing\footnotemark} & How individuals feel about and evaluate their lives. & \\
\quad \textit{\textbf{Affective Wellbeing}} & The presence of positive emotions (e.g. happiness, contentment, joy), and the relative absence of negative emotions (e.g. sadness, anger, stress). & Self-report scales of current or recent affect; e.g. Positive and Negative Affect Schedule (PANAS) \citep{watson_1988_development}, Day Reconstruction Method (DRM) affect ratings \citep{kahneman_survey_2004}. \\
\quad \begin{tabular}[t]{@{}l@{}}
\textbf{\textit{Evaluative Wellbeing}}
\end{tabular}
& A reflective cognitive evaluation of one's life overall, typically derived from comparing current circumstances to self-imposed standards or ideals. & Agreement with statements like ``I am satisfied with my life'' or ``In most ways my life is close to my ideal'' \citep{diener_satisfaction_1985}. \\
\midrule[0.1pt]
\textbf{Eudaimonic Wellbeing} & The extent to which individuals realize their potential and optimal functioning, understood in terms of meaning, purpose, self-actualization, and/or autonomy. & Self-report along dimensions including: purpose in life, autonomy, self-acceptance, and personal growth \citep{ryff_happiness_1989, ryff_structure_1995}.\\
\midrule[0.1pt]
\textbf{Psychological Health} & The absence of psychopathology and mental distress. & Standard screening questionnaires, e.g., GHQ-12 \citep{goldberg_users_1991} or SF-36 mental health scale \citep{ware_mos_1992}; clinically diagnosed mental disorders (e.g., anxiety, depression).\\
\textbf{Physical Health} & Bodily health, including physical functioning, absence of disease, and lower mortality. & Standardized health-related quality of life assessments; diagnosed health conditions; life expectancy. \\
\textbf{Cognitive Health} & The capacity to perform core mental processes, such as memory, learning, and executive functioning. & Performance on specific cognitive tasks (e.g. word recall, verbal fluency, digit span). \\
\bottomrule
\end{tabular}
\caption{Dimensions of Wellbeing}
\label{tab:wellbeing_framework}
\end{table*}

\subsection{Related work and our contributions}

The question of how AI-driven automation will affect human wellbeing sits at the intersection of several literatures, none of which, on its own, provides an integrated empirical basis for anticipating these impacts.

A first strand of research focuses on \textit{AI labor economics}---forecasting which jobs are susceptible to automation and estimating the scale of potential displacement \citep[e.g.,][]{Frey2017-in, Brynjolfsson2016-kf, acemoglu_simple_2025}. This literature has grown rapidly, producing increasingly sophisticated models and empirical evaluations of task exposure and labor market adjustment \citep[e.g.,][]{autor_expertise_2025, dellacqua_navigating_2026}. However, these analyses are primarily concerned with economic output and employment levels, and rarely address the \emph{wellbeing} consequences of the transitions they forecast.

A second strand addresses \textit{policies for AI economic transitions}. Proposals range from financial transfers like universal basic income (UBI) and universal basic capital (UBC) \citep{korinek_preparing_2024, Freeman2015-sc}, to state-provisioned employment such as job guarantee programs \citep{tcherneva_job_2018}, and retraining funds \citep{susskind_world_2020}. These proposals represent important contributions to the policy landscape and frequently recognize the non-monetary benefits of work. However, while they draw on valuable sociological and psychological insights, they typically focus on isolated examples. Our paper builds a more comprehensive foundation for these debates by providing the first comprehensive, cross-population empirical synthesis of how real populations fare without traditional employment. Our review highlights cross-cutting empirical evidence that financial transfers alone (as in UBI or UBC) may be insufficient to sustain wellbeing, not just because work provides significant non-financial benefits (\S\ref{sec:benefits_of_work}), but also because populations dependent on material redistribution risk the erosion of agency (\S\ref{sec:financial_dependence}). Furthermore, we find that populations that do not \emph{choose} their work status have significantly lower wellbeing than those who do (\S\ref{sec:voluntary_vs_involuntary_retirement}--\ref{sec:stay_at_home_mothers})---this is a dimension not normally considered in policy discussions. Finally, while job guarantee discussions extensively invoke historical, Western work-relief programs like the New Deal, our synthesis highlights highly relevant modern examples that are less discussed, such as state-provisioned employment in the Gulf States (\S\ref{sec:state_employment}).

A third strand, primarily in \textit{philosophy and ethics}, examines questions of meaning, purpose, and flourishing in relation to work. Some authors have argued that AI risks creating an ``achievement gap'' by displacing the challenges and tasks that make work fulfilling \citep{nyholm_meaning_2023, Danaher2019-ff}, while others have analyzed the non-monetary goods that work provides, e.g. community, social contribution, and excellence \citep{Gheaus2016-ht}. This philosophical work provides valuable conceptual scaffolding for thinking about work and human flourishing, but it typically does not engage in depth with the empirical evidence on how actual non-working populations experience these dimensions of wellbeing.

Finally, there is a rich body of \textit{empirical research on work and wellbeing} in psychology, sociology, and economics. Within specific populations, sophisticated meta-analyses and reviews have documented the negative effects of unemployment \citep[e.g.,][]{paul_unemployment_2009, gedikli_relationship_2023}, the wellbeing dynamics of retirement \citep{wang_profiling_2007, chen_review_2025}, and the theoretical frameworks linking work to psychological functioning \citep{jahoda_work_1981,fryer_employment_1986}. Our paper draws extensively from this literature as its primary evidence base. However, existing reviews are largely siloed by population. To our knowledge, no prior work has synthesized evidence \textit{across} multiple non-working populations---comparing the unemployed, retirees, lottery winners, dependent spouses, and others---to identify the common factors that shape wellbeing without work. Still less has such a synthesis been organized around AI-driven automation scenarios.

This paper aims to fill these gaps. By providing the first cross-population review organized around AI futures, we identify three key factors that mediate the relationship between work status and wellbeing: (1) \textit{agency and choice}---whether the exit from work is voluntary or involuntary, as well as long-term agency; (2) the availability of \textit{alternative sources of work's latent benefits}---such as volunteering, hobbies, or state-provisioned employment; and (3) \textit{social and systemic context}---including cultural norms around work and the robustness of social safety nets. We draw on these three factors to derive specific implications for different AI automation scenarios, connecting the empirical evidence to concrete policy considerations.

\footnotetext{We use ``Subjective Wellbeing'' in the narrower sense common in the empirical psychology literature \citep[e.g.,][]{Diener1984-cx, gedikli_relationship_2023}, referring to hedonic experience and evaluative life satisfaction. This is also how the term is generally used in the studies reviewed here. Some frameworks \citep[e.g.,][]{OECD-2013-tn} use the term more broadly to also encompass eudaimonic wellbeing, but we present eudaimonic wellbeing separately here to be consistent with the psychological literature.}

\subsection{The populations studied}

This comparative review draws from studies across OECD countries, China, India, and Gulf states. Table \ref{table:populations} provides an overview of the populations surveyed in this work. The outcomes for these populations may bear upon the pitfalls and possibility of wellbeing without work. When juxtaposed against possible AI futures, these groups may help to illuminate different aspects of the choices we now face.

These populations are deliberately heterogeneous, and provide contrasting cases whose differences help isolate the factors that shape wellbeing without work. Although no group is a direct analogue for any specific future, each group may be particularly informative for understanding specific aspects of these scenarios. Furthermore, where findings converge across different contexts (e.g. as they do for the importance of agency and choice), we can be particularly confident that the shared moderating factors are robust.

\subsection{Measures of wellbeing}

Wellbeing can be understood and measured in a range of ways. Table \ref{tab:wellbeing_framework} summarizes the major dimensions of wellbeing and measures reviewed in this survey. Subjective wellbeing, eudaimonic wellbeing, psychological health, cognitive health and physical health feature prominently in these studies. These wellbeing dimensions and measures are commonly used in psychology and social sciences. Rather than forming strict, separate categories, these dimensions of wellbeing frequently overlap. At the same time, they often exhibit distinct dynamics and respond differently to external factors \citep[e.g.,][]{eid_global_2004,schimmack_influence_2008,wiest_subjective_2011, kahneman_high_2010}.

\subsection{Scope and approach}
\label{sec:scope_and_approach}

This paper offers a comparative, cross-disciplinary synthesis of empirical evidence on non-working populations, drawing from psychology, economics, and sociology. Our research question---how existing evidence on non-working populations may inform us about wellbeing in AI futures---spans disparate fields, populations, and methodologies, which cannot be meaningfully aggregated through a single systematic protocol. Instead, we adopt the approach common to review articles in the social sciences \citep[e.g.,][]{Baumeister1997-vs, Sukhera2022-lj}, where the goal is to build an integrative empirical synthesis rather than to exhaustively answer a narrow empirical question.

Literature was identified through targeted searches across academic databases (including OpenAlex, Google Scholar, PsycINFO, EconLit, and PubMed), and supplemented by backward and forward citation tracing of seminal studies and meta-analyses. Each search round was followed by a gap analysis that identified underrepresented populations, outcome types, or methodological approaches, which informed subsequent targeted searches across multiple cycles. We prioritized meta-analyses, longitudinal panel studies, and studies with stronger causal identification strategies---natural experiments (e.g. mass layoffs, lottery wins, policy discontinuities), randomized controlled trials---while also noting findings from cross-sectional studies where relevant. Appendix Table \ref{table:studies_and_techniques} provides brief descriptions of the study types referenced throughout. The selection and interpretation of studies reflects the authors' judgment; we do not claim exhaustive coverage of any single sub-question, and we view this work as a potential foundation for future reviews that may be more focused on specific sub-topics.


\section{The costs of unemployment}
\label{sec:unemployment_costs}

When it comes to understanding the impact of automation and potential loss of work on human wellbeing, the experience of unemployed people is critical. Specifically, we refer to the population normally defined by economists as ``unemployed'' (i.e. individuals who are jobless, actively seeking work, and available for work). There have been extensive, validated, cross-country studies exploring the impact of unemployment and job loss on mental and physical wellbeing. This section reviews the evidence and notes that unemployment is generally associated with negative trends in wellbeing. In the context of conversations about AI transformation, a key question is how such effects can be mitigated or avoided.

\subsection{The negative effect of unemployment on wellbeing}
\label{sec:unemployment_wellbeing}

Unemployment typically causes a significant decline in wellbeing. Longitudinal studies and meta-analyses from around the world consistently indicate that unemployment causally decreases both mental health and evaluative wellbeing \citep{jefferis_associations_2011, gedikli_relationship_2023}. Natural experiments involving mass layoffs and plant closures provide strong evidence for a causal link in those contexts \citep{lawes_impact_2023, riumallo-herl_job_2014}.
These negative effects extend beyond just the effects of decreased income \citep{flatau_mental_2000}. They also worsen over time, the longer an individual remains unemployed \citep{gedikli_relationship_2023}.

While there is clear evidence for negative impacts on evaluative wellbeing (measured in terms of life satisfaction), the impact of unemployment on daily \textit{affective} wellbeing (day-to-day emotional experience) may be more nuanced. Working is often considered unenjoyable\footnote{Though not uniformly \citep{Strauss2024-fl,Strauss2024-om}}, and unemployed people can dedicate more time to leisure instead. Some research suggests that this ``time-composition effect'' leads the average self-reported affective wellbeing of employed and unemployed people to be largely the same, even while the unemployed experience significantly decreased life satisfaction \citep{knabe_dissatisfied_2010}.

People who have experienced unemployment may even exhibit long-term psychological ``scarring'' that persists after individuals regain employment \citep{lucas_unemployment_2004, luhmann_subjective_2012}. Scarring may be particularly severe for individuals who experience unemployment in their youth \citep{strandh_unemployment_2014}. While a recent U.S. study used updated methods to challenge the existence of scarring effects \citep{rauf_getting_2021}, a later German study using the same methods continued to find long-lasting negative effects that persisted for at least five years after re-employment \citep{eberl_subjective_2023}.

Unemployment tends to have a similarly negative and long-lasting effect on physical health. In a study of plant closures, job loss was found to increase the risk of overall mortality, of suicide and suicide attempts, and of death and hospitalization due to traffic accidents, alcohol-related disease, and mental illness \citep{browning_effect_2012}. Strikingly, the authors found that these negative effects on mortality tended to be very large in the year of displacement  (79\% higher), but still persisted at a lower level even 20 years later (11\% higher). Moreover, these mortality effects were observed in Denmark, a country with a robust welfare system---we might expect even stronger effects in countries with weaker social safety nets, all else equal (see further discussion of social safety nets in \S\ref{sec:safety_nets}). Additionally, job loss has been found to trigger a measurable increase in physiological stress markers, like cortisol or interleukin-6 \citep{hintikka_unemployment_2009, lawes_unemployment_2022}.

\subsection{Implications for AI futures}

Overall, this evidence suggests that automation could pose a serious challenge for wellbeing, if it leads to job loss under conditions similar to those experienced by unemployed people today. If AI displacement frees workers from unenjoyable tasks, the ``time-composition effect'' suggests that day-to-day \emph{affective} wellbeing may not suffer, as people substitute work with leisure. However, the evidence points to a potential crisis for \emph{evaluative} wellbeing, as well as for mental and physical health. 

The negative effects on mental health appear to be substantial and long-lasting, as evidenced by the literature on ``psychological scarring''. Moreover, unlike cyclical unemployment, where displaced workers can typically expect to return to similar roles, AI-driven automation may permanently eliminate certain job categories \citep{Frey2017-in, susskind_world_2020}. This would make transitioning back into the workforce much harder for those with specialized skills, increasing the risk of long-term (rather than temporary) job loss. The evidence reviewed above suggests this could amplify the psychological toll---prolonged unemployment is associated with worsening wellbeing over time. Similarly, the mortality evidence from plant closures indicates a physical health toll from job displacement that can persist for a long duration, even in countries with strong welfare systems. Because this psychological and physical toll begins immediately, any AI-driven labor market upheaval may require interventions that are both swift and structurally enduring.

Sections \ref{sec:benefits_of_work}--\ref{sec:preference_for_work} add further definition to this picture, exploring the particular benefits of work that may be lost, and the extent to which people choose to work even when it is not financially necessary. The conditions experienced by the unemployed may represent a worst case, as they are actively seeking work, cannot find it, and unemployment is \textit{unevenly} experienced across the population. Sections \ref{sec:agency_and_choice}--\ref{sec:societal_systemic} explore alternate scenarios and factors that may substantially moderate these effects.


\section{The benefits of work beyond income}
\label{sec:benefits_of_work}

To better understand why unemployment is causally associated with loss of wellbeing, it is important to consider the different benefits that work provides. These include direct benefits via income, as well as indirect benefits such as status and time structure. What does the evidence tell us about these benefits, and their associations with wellbeing for working and non-working individuals?

\subsection{Latent and manifest benefits}
\label{sec:latent_and_manifest}

In a seminal theoretical paper, \citet{jahoda_work_1981} proposed that while the \textit{``manifest''} or intended purpose of employment is to earn a living, work also provides five \textit{``latent''} or unintended psychological benefits: time structure, social contact, status and identity, regular activity, and collective purpose understood as shared goals that transcend an individual's own. Jahoda posited the loss of these latent benefits during unemployment as the primary driver of psychological distress. In contrast, \citet{fryer_employment_1986} argued that the main negative consequence of unemployment arises from the loss of the \textit{``manifest''} benefits---most importantly, income. He posited that the profound financial strain restricts a person's agency and ability to plan for the future, and that this is the primary cause of negative psychological effects.

The evidence seems to show that both latent and manifest benefits are necessary for explaining the positive contributions of work to wellbeing, and, conversely, the negative impacts of unemployment. \citet{creed_relative_2001} directly compared these two theories in a cross-sectional study of unemployed individuals in Australia, showing that both models are necessary for a complete explanation of the subjects' psychological wellbeing. They found that financial strain was the largest predictor of wellbeing, while some latent benefits were also significant predictors---status, time structure, and collective purpose (but not social contact or regular activity).\footnote{Social connectedness is often thought to be one of the major benefits of work. However, the evidence on this point is nuanced. Cross-sectional, cross-national studies across Europe show that unemployment does not necessarily reduce social contact with friends and family, and in some cases the unemployed are \textit{more} likely to be in the most highly sociable category \citep{gallie_unemployment_1999, rozer_does_2020}. Where unemployment does significantly reduce social participation (in clubs and organizations and financially costly social outings), this seems driven in large part by poverty risk and financial hardship, acting as a downstream effect of job loss rather than the loss of the workplace as a social venue \citep{dieckhoff_unemployed_2015}. Similarly, other research suggests that the unemployed socialize in highly segregated networks with other jobless peers \citep{gallie_unemployment_1999}, which may be driven by a desire to intentionally isolate themselves to avoid social stigma \citep{peterie_social_2019}.}

\subsection{Purpose and meaning}
\label{sec:eudaimonic_benefit}

A sense of meaning and purpose may serve as yet another benefit of work---and that work contributes to a kind of eudaimonic wellbeing (which is broader than Jahoda's proposed benefit of ``collective purpose'', since it is not necessarily tied to goals that are shared with others). In particular, meta-analyses have shown a negative association between retirement and eudaimonic wellbeing \citep{pinquart_creating_2002}. Still, these effects are small and the relationship here is complex---for those in lower socioeconomic status or physically demanding low-autonomy jobs, retirement often sparks an \textit{increase} in purpose \citep{yemiscigil_effects_2021}. Furthermore, certain kinds of bridge employment or volunteering can help mitigate this loss of purpose and meaning (see Section \ref{sec:alternate_activities} below).

\subsection{Cognitive health}
\label{sec:cognitive_benefit}

Evidence from retirees suggests that cognitive health is another important latent benefit of work. Retirement is causally linked to decreases in cognitive abilities, consistent with a ``use it or lose it'' perspective. For example \citet{rohwedder_mental_2010} leveraged variation in retirement policies across countries, finding causal evidence that early retirement leads to significant drops in cognitive ability. This effect was subsequently confirmed within the US and the UK \citep{bonsang_does_2012,xue_effect_2018}.

\subsection{Implications for AI futures}

The evidence shows that employment provides a range of significant benefits beyond income. For example, paid work can provide individuals with status, time structure, and purpose. There is also strong evidence that work helps adults maintain cognitive health, with retirement being causally associated with cognitive decline. Contrary to popular belief, social connectedness does not appear to be an independent benefit of work: where unemployment does reduce social participation, this appears to be driven largely by the financial deprivation and loss of social standing that accompany joblessness, rather than by the loss of the workplace as a social venue.

Given this broad basket of non-financial benefits from work, any post-labor society must consider how to replace or sustain them outside of traditional employment. Furthermore, policy interventions focusing purely on wealth distribution (e.g. UBI or social dividends) may fail to address these broader needs, because they only replace the \emph{manifest} benefits of work.

One option would be to guarantee jobs, or to ensure that there is work even if the amount of time invested is dramatically reduced (e.g. \S\ref{sec:state_employment}). Another option would involve discovering or investing in new activities that combine similar properties---perhaps to a higher degree than in the modern workplace, where many of these effects are epiphenomenal rather than intentional. Fortunately, we shall see (\S\ref{sec:alternate_activities}) that there are a number of activities that are known to mimic the benefits of paid work, such as volunteering, which could be productively tapped into across a range of economic futures.


\section{The preference for work}
\label{sec:preference_for_work}

Given the psychological and non-financial benefits of work, it is perhaps unsurprising that many choose to work even when it may not be financially necessary. This trend is most clear for lottery winners, as detailed in the case study below.

\subsection{Lottery winners and employment}

Evidence from lottery winners suggests that while wealth buys leisure, it rarely results in full exit from the workforce. In multiple studies of U.S. and Swedish lottery winners, the vast majority of people choose to continue working after their windfall (consistently 85\%--89\% across studies) \citep{furaker_gambling_2012, arvey_work_2004, kaplan_lottery_1987}. Such consistency is striking given differing work cultures in the U.S. and Sweden, the expanse of the data across multiple decades, and the fact that the average lottery prize ranged from just under \$400,000 in the Swedish study to an average of \$7.2 million in one of the U.S. studies. Moreover, the U.S. study showed that, among those who did choose to continue working, the average winnings were still very high (\$5.2 million).\footnote{All amounts converted to 2026 dollars.}

The inclination to stay employed was stronger when individuals had higher ``work centrality'' (the degree to which work was central to their identity) and when they were already in psychologically rewarding roles \citep{arvey_work_2004,kaplan_lottery_1987}. Older winners were more likely to retire \citep{kaplan_lottery_1987}.

Instead of abandoning work entirely, lottery winners typically use their wealth to purchase \emph{flexibility}. Swedish data reveals that many winners took unpaid full-time leave for a few weeks at a time, and others reduced their working hours \citep{furaker_gambling_2012}. Underscoring this point, lottery winners spend approximately 11\% of their unearned income on purchasing leisure time over their lifetime \citep{cesarini_effect_2017}.


\section[Decoupling wellbeing from traditional market-driven employment]{\fontsize{13.5}{16}\selectfont\textls[-45]{Decoupling wellbeing from traditional market-driven employment}}
\label{sec:decoupling}

In \S\ref{sec:unemployment_costs}, we documented the negative effects of unemployment on wellbeing. In this section, we look at the contrast case and ask: are there populations who \textit{do} have high wellbeing without traditional full-time labor? We survey the evidence on retirees, and also the literature on alternate activities that may replicate the benefits of work. We also review the evidence regarding state-provisioned employment.

\subsection{Sustained wellbeing in retirement}
\label{sec:retiree_wellbeing}

Retirees often report high levels of subjective wellbeing---or even increases in life satisfaction---after exiting the workforce. In reviews across multiple large-scale U.S. panel datasets \citep{chen_review_2025, szinovacz_m_e_contexts_2003}, over 90\% of retirees are satisfied with their lives, a trend that has remained stable over decades. In fact, for adults aged 60 or over (a population that includes both retirees and non-retirees), life satisfaction scores approximately match the national average, and this group enjoys relatively low rates of depression and anxiety.\footnote{Six U.S. datasets consistently indicate that adults aged 60+ report life satisfaction at 7 or 8 on a 10-point scale. For comparison, average life satisfaction for Americans was 7.7 \citep{radcliff_politics_2001}, and for US college students it was 6.3 when normalized to a 10-point scale \citep{pavot_satisfaction_2008}.} This may be especially surprising given the general decline in health at this stage in life, the loss of work, and decreases in income.

Longitudinal tracking demonstrates that these effective psychological adjustment patterns can be observed specifically in the transition to retirement. \citet{wang_profiling_2007} found that the majority of retirees ($\sim$70\%) exhibit a ``maintaining'' pattern, with minimal change in wellbeing through the transition. Others followed a U-shaped pattern with an initial decline followed by recovery ($\sim$25\% of retirees), or a pattern of linear improvement, often among those retiring from stressful or unhappy circumstances ($\sim$5\% of retirees). Notably, the study did not identify a subgroup of retirees who experienced a long-term, permanent decrease in wellbeing.\footnote{However, the samples only included retirees who survived for at least $\sim$8 years into retirement, such that survivor bias might partially influence these findings.}

Human adaptability and resilience are well-documented phenomena in the wellbeing literature, and retirees often demonstrate an effective capacity to navigate the loss of work, income, and physical health. However, these aggregate positive trends mask important heterogeneity and trade-offs. Most notably, the benefits of retirement are strongly mediated by agency: as we explore in \S\ref{sec:agency_and_choice}, the 20-30\% of individuals who are \textit{forced} into retirement do not experience these positive trends, suffering declines in wellbeing more akin to unemployment. Furthermore, the positive effects on subjective wellbeing must be weighed against potential costs, such as the established causal link between retirement and cognitive decline (\S \ref{sec:cognitive_benefit}).

Taking together the evidence on retirees and the unemployed, we see that ``not working'' is not a singular state---the effects on wellbeing vary substantially depending on context. \S\ref{sec:agency_and_choice} and \ref{sec:societal_systemic} provide deeper analysis of the factors driving such divergent outcomes, such as whether the exit from work is voluntary, and the role of social norms.

\subsection{Alternate activities as sources of wellbeing benefits}
\label{sec:alternate_activities}

While innate psychological adaptation may help individuals adapt to the loss of work, wellbeing in retirement is also bolstered by the specific activities undertaken by retirees. These activities can help to replace the structure, purpose, and sense of identity previously provided by employment.

\emph{Volunteering} serves as a protective factor for older adults, with formal volunteering especially offering structure and social identity that mimics the work role. Volunteering can significantly improve life satisfaction, purpose in life, cognitive health, physical health, and psychosocial wellbeing \citep{greenfield_formal_2004, kim_volunteering_2020, han_helping_2025}. Notably, the threshold for these benefits is low---benefits are observable at just 1 hour of volunteering per week, with marginal gains beyond 2 hours per week \citep{jongenelis_longitudinal_2022,lum_effects_2005, kim_volunteering_2020}. These benefits are stronger with a greater variety of activities and also when the individual feels appreciated \citep{matthews_impact_2020, mcmunn_participation_2009}.\footnote{Overall, these studies among older adults are largely consistent with studies demonstrating the beneficial effects of volunteering on mental health and mortality among the wider population, e.g. as reviewed by \citet{Meier2007-wp,jenkinson_is_2013,Binder2013-eg}.}

\emph{Hobbies} can also support mental health and provide protection against physical and cognitive decline. Hobbies are associated with healthier aging especially when they involve physical activities \citep{mak_hobby_2023, tomioka_relationship_2016}. Leisure activities (e.g. clubs, hobbies and cooking) and physical activities are both linked to improved mental health and lower depression in old age \citep{vo_mental_2023, bone_engagement_2022}. And arts engagement is associated with benefits for cognitive functions \citep{bone_participatory_2024}.

Finally, \emph{caregiving} is important but yields outcomes that are more mixed, depending largely on the intensity of the caregiving role. It can be both a source of deep satisfaction and stress for older and retired adults. Occasional caregiving, for example babysitting grandchildren, is generally associated with better mental health \citep{notter_grandchild_2022}. However, intensive caregiving, such as being the primary caretakers of grandchildren, can increase stress and decrease wellbeing overall \citep{waldrop_grandparent_2001, notter_grandchild_2022, szinovacz_predictors_2005}. This strain is more pronounced in grandfathers than grandmothers, across both American and Chinese populations, perhaps due to differing socialization between men and women into caregiving roles \citep{waldrop_grandparent_2001, notter_grandchild_2022, szinovacz_predictors_2005}.

\subsection{State-guaranteed employment}
\label{sec:state_employment}

Gulf State nationals demonstrate how state-guaranteed employment---which was widespread before the oil price crash of 2014---can also be associated with high levels of subjective wellbeing. In Arab countries, public sector workers report higher levels of happiness and life satisfaction than private sector employees. This is attributed to the sector's historical role as a pillar of the welfare state, offering job security, wage premiums, and lighter workloads \citep{mikhaeil_well-being_2025}.

These roles are so desirable that youth unemployment, which was historically high in the region, is often the product of voluntary choice---with young citizens preferring to wait for openings in the public or semi-public sectors, rather than accepting private sector employment \citep{hani_understanding_2021}. 

Moreover, state-provisioned employment may be beneficial for physical wellbeing and future economic outcomes. The Civilian Conservation Corps was a U.S. work relief program to combat high unemployment during the Great Depression, employing young men in environmental conservation. \citet{aizer_youth_2020} found that program participation increased lifetime longevity by approximately one year, increased physical height, and resulted in 4.6\% higher lifetime earnings.\footnote{Notably, there were no short-run changes in wages or labor force participation---this emphasizes the importance of looking at long-term effects, even when there are no apparent short-term changes.} The same study included a randomized controlled experiment with Job Corps, a modern version of the program from the 1990s, and found similar results.

\subsection{Implications for AI futures}

In Section \ref{sec:benefits_of_work}, we established that a post-labor society would need to replace the non-financial benefits of work. The evidence surveyed here indicates that it is indeed possible to decouple such benefits from traditional employment. Among retirees, alternate activities such as volunteering and hobbies are associated with recovering some of the wellbeing and non-financial benefits that work provides. This may help to explain the high life satisfaction and psychological adaptability among retirees. Furthermore, the thresholds for achieving such benefits can be remarkably low (e.g. \mbox{1-2} hours of volunteering a week in some studies, at least for older adults), indicating that people may not require a volume of purposeful, structured activity equivalent to a full-time job. Policies in a post-labor world should not focus solely on replacing income, but also on actively cultivating and subsidizing avenues for such activities.

Alternately, in scenarios where the free market fails to provide sufficient employment, state-provisioned or guaranteed employment may present a viable alternative for populations who still desire or require the formal structure of a job. This is evidenced by the robust wellbeing associated with public sector work in the Gulf States or historical work-relief programs in the U.S.

However, we should be circumspect in the implications that we draw, for two reasons. First, the high wellbeing reported in the Gulf States may be inflated by the unusual wage premiums, benefits, and lighter workloads of those specific public sector jobs. (Although the desirability of lighter-workload public sector jobs in Gulf States is consistent with the abovementioned evidence on volunteering, with lower volumes of work potentially being sufficient to sustain wellbeing.) Second, relying on state-provisioned employment may create systemic dependencies that can ultimately erode societal agency, a risk we explore further in Section \ref{sec:agency_and_choice}. Nonetheless, these examples are encouraging and demonstrate the \emph{compatibility} of state-guaranteed employment with high wellbeing.

In practice, there is unlikely to be a single, monolithic replacement for market-driven employment. Individuals vary in how they derive time structure, purpose, and social identity---some may thrive in more self-directed roles while some may prefer more formal volunteering, while others may prefer the structure of a traditional job, even if that job is state-provisioned. If the landscape transforms, a portfolio of options should be available, enabling individuals to secure the latent wellbeing benefits of work in ways that best suit their preferences and circumstances. This also helps to ensure that individuals retain agency and choice---we further discuss the importance of these factors in the next section.


\section{The importance of agency and choice}
\label{sec:agency_and_choice}

The discrepancy in wellbeing between the unemployed and retirees is striking. How might this discrepancy be explained? We propose that one major factor is the role played by \textit{agency and choice}---unemployment is an involuntary state by definition, whereas retirement is often voluntary.

In the next section, we see that differences in agency and choice lead to divergent wellbeing outcomes even within the retiree population---where \textit{involuntary} retirement looks more similar to unemployment---and in other populations such as stay-at-home mothers, too. 

Furthermore, agency is not just about the freedom to choose \textit{whether} to work, but also maintaining autonomy and leverage no matter the source of one's material or financial needs. We examine financially dependent populations, such as financially dependent spouses, to understand such settings. These populations may inform expectations about potential AI futures characterized by human dependency on AI, the state, or other forms of centralized power.

\subsection{Voluntary vs involuntary retirement}
\label{sec:voluntary_vs_involuntary_retirement}

While retirement is usually framed as a choice, a significant minority of workers (20-30\%) perceive their retirement as \textit{forced}, e.g. due to health limitations, job displacement, or caregiving obligations \citep{szinovacz_predictors_2005}. Across countries, voluntary retirees consistently report significantly higher physical and mental wellbeing than involuntary retirees, even when controlling for factors such as health, income, job characteristics, and life events \citep{calvo_gradual_2009, bender_analysis_2012, mosca_impact_2014}. 

The case for the importance of choice and agency is further strengthened when considering different motivations for retirement, as proxies for agency in the decision to retire. When retirement is driven by internal ``pull'' factors (e.g. the desire to pursue personal interests, or spend time with family)---functioning as a form of self-determination---we do not observe the same steep declines in mental health associated with ``push'' factors driven by external factors (e.g. pressure from employers, layoffs, or poor health) \citep{negrini_push_2013,vo_mental_2023}.

\subsection{Stay-at-home mothers}
\label{sec:stay_at_home_mothers}

For mothers, similarly, the relationship between wellbeing and work is more strongly influenced by \textit{personal preferences} than employment status \textit{per se}. One study found that mothers' wellbeing was most strongly determined by the match between their employment status and their wish to work  \citep{ciciolla_what_2017}. It found that mothers who stayed at home but desired to work reported the poorest adjustment---worse even than those who worked primarily for financial necessity but preferred to be at home.

\subsection{Financial dependence}
\label{sec:financial_dependence}

Financial resources provide one of the most important means for empowerment in modern societies. Unsurprisingly, therefore, financial dependence is associated with disempowerment and loss of agency. This is evident in marriages, where financial independence enables women to escape unhappy marriages (without increasing divorce rates for women who are satisfied with their marriages) \citep{sayer_she_2011}. This study found no comparable impact of women's employment on men's decisions to exit a marriage, but married men do experience higher psychological discomfort when they are fully economically dependent on their partners than when they have roughly equal contributions to household income \citep{syrda_spousal_2020}.

The risks of financial dependence can compound over time, as evidenced by ``gray divorces'' (occurring after age 50). Because women have historically sacrificed labor force participation for child-rearing, they suffer from long-term resume and income gaps, even if they are working at the time of the divorce. Following a divorce, women experience on average a  45\% decline in their standard of living, compared to a 21\% decline for men \citep{lin_economic_2020}. This demonstrates that long-term financial dependence leaves individuals highly vulnerable if the relationship dissolves.

The vulnerabilities associated with dependence can also manifest at a societal scale. In the Gulf States, for example, the historical reliance on state-provisioned public sector employment has created systemic dependencies. While this model has generally yielded high wellbeing (as reviewed above in \S\ref{sec:state_employment}), economic shifts following the 2010s oil price crash highlight the risks of this social contract. Efforts to diversify economies and shift national labor to the private sector have encountered friction due to historic labor market structures, such as high reservation wages and the ongoing need for broader human capital development \citep{kayan-fadlelmula_systematic_2022}. And when the state reduced subsidies on necessities, prompting public pushback, authorities frequently responded by restricting civic space and increasing state repression \citep{krane_subsidy_2019}.
Adjusting to a changing state economic role can therefore be challenging for both individuals and societies.

\subsection{Implications for AI futures}
\label{sec:agency_implications}

The evidence considered above suggests that a critical determinant of wellbeing is not just the status of one's employment, but rather the \textit{agency and choice} associated with that status. There is a profound divergence in outcomes for individuals in voluntary vs involuntary work states. Furthermore, material or financial dependence can itself lead to a worrisome erosion of agency. Such findings are potentially highly consequential when it comes to understanding possible AI futures. If AI transforms the labor landscape, a primary objective should be to safeguard individual agency both \textit{during} and \textit{after} any period of transition.

During transitions, we find that individuals who exit the workforce voluntarily may have much better wellbeing outcomes than those who exit involuntarily, even if material circumstances are held constant for both groups. Many policy discussions about AI futures focus on the availability of work or the availability of resources. However, these discussions overlook the fact that it matters deeply whether people get to \emph{choose} their path. When transitions are imposed without consent, the resulting loss of agency severely diminishes wellbeing. This implies the importance of providing options and autonomy in navigating potential AI-driven economic transformations.

In the long run, we should also be wary of scenarios characterized by systemic dependency, as previously raised by some others \citep{kulveit_gradual_2025}. The evidence on financially dependent spouses and the Gulf States illustrate a few different risks---populations that are systemically dependent may lose their ability to navigate changing circumstances, or to exit unfavorable arrangements. 

Overall, policies or societal structures that only address material or other basic needs (without addressing agency) are likely insufficient to ensure wellbeing.


\section{Social and system-level factors}
\label{sec:societal_systemic}

This section highlights important social and cultural effects moderating the relationship between work and wellbeing, as well as other systemic factors. The relationship between work and wellbeing is strongly affected by norms and societal expectations (\S\ref{sec:social_norms}), the local prevalence of unemployment (\S\ref{sec:others_unemployment}) and the existence of social safety nets (\S\ref{sec:safety_nets}).

\subsection{Social norms around work}
\label{sec:social_norms}

The psychological cost of unemployment is heavily influenced by societal expectations. A meta-analysis across 29 studies found that a country's ``norm to work'' was significantly correlated with the negative wellbeing effects of unemployment \citep{gedikli_relationship_2023}.
A cross-national European study also found that the wellbeing cost of unemployment is significantly higher in more individualistic countries \citep{mikucka_does_2014}.

The phenomenon of ``retiring from unemployment'' is a striking illustration of the power of norms. Specifically, the phenomenon encompasses individuals who were initially unemployed and looking for work, but who then transitioned into retirement after reaching eligible age. In a German study, those who went through this transition experienced a significant increase in life satisfaction, \textit{even though they experienced essentially no change in income or daily routine} \citep{hetschko_changing_2014}. A wider European study similarly found a 23\% decrease in depression when individuals pass the Early Retirement Age in their country, and again this was true despite no significant increase in government transfer or social activities after this transition \citep{van_de_kraats_why_2024}. Thus, there seems to be a wellbeing penalty that arises from the status of being an ``able-bodied'' person without a job, which is lifted once a person enters an age where it is acceptable to be unemployed.\footnote{These studies both took place in European countries. In countries with weaker social safety nets (like the U.S.), the negative wellbeing effects of unemployment might be driven more strongly by financial factors (e.g. loss of housing or health benefits) than the psychological loss of social identity.} In many modern societies, retirement is often considered an ``earned" unemployment or as reward for having ``paid one's dues" \citep{kohli_retirement_1987,jonson_are_2007,smith_revealing_2012}.

The importance of socio-cultural norms is also evidenced by comparing wellbeing outcomes for men vs. women. Across many studies and settings, the negative psychological effects resulting from loss of work are exhibited more strongly by men than by women \citep{gedikli_relationship_2023, paul_unemployment_2009, clark_boon_2010, hetschko_changing_2014}, presumably because of stronger social pressure to adhere to work-related norms, and less socialization into meaningful roles outside the workplace such as caregiving. This discrepancy extends even to retirement-related cognitive decline, which is worse for men than for women \citep{atalay_effect_2019}. A German study underlined the importance of role-based expectations in the context of relationships: both men and women experience strong negative drops in life satisfaction when they become unemployed, but living in a partnership exacerbated this effect for men, while weakening it for women \citep{knabe_partnership_2016}.

\subsection{Others' unemployment}
\label{sec:others_unemployment}

We can further investigate the idea that wellbeing effects are impacted by socially moderated expectations, by observing how ``work norms'' can also shift based on local conditions. A British study examined the phenomenon of ``unemployment as a social norm'', showing that unemployment has a less negative effect on psychological wellbeing when ``relevant others'' are also unemployed \citep{clark_boon_2010}. The salient reference groups included: an individual's partner, all other adults in the household, or people of the same sex living in the same region.\footnote{However, the negative effects on \textit{mortality} are more pronounced when local unemployment is high \citep{browning_effect_2012}.} Similarly, ``joint'' retirement, where partners retire simultaneously, is generally associated with positive outcomes, whereas asynchronous retirement can create friction and lower marital satisfaction \citep{moen_couples_2001}.

These results suggest that individuals evaluate their employment status against ``local norms'' established by their peers (whether within the household or across a region). The psychologically protective effects of high regional unemployment are particularly notable, given that re-employment is actually more difficult in such settings.

For people who are currently \textit{employed}, on the other hand, higher unemployment among relevant others can sometimes \textit{decrease} their wellbeing \citep{clark_boon_2010, borra_wellbeing_2016}. This may be caused by feelings of job insecurity, feelings of guilt, or individuals staying in unsatisfactory jobs that they would otherwise have quit in better labor market conditions. At the same time, there may simultaneously be a positive ``comparison'' effect where individuals feel better off than their peers, increasing their job satisfaction \citep{borra_wellbeing_2016}.

\subsection{Social safety nets}
\label{sec:safety_nets}

Finally, while cultural norms heavily influence the subjective experience of job loss, institutional safety nets play an important role shaping its material impacts. Alternatives for sustaining a living can extend beyond the welfare state to include families, local communities, and voluntary associations. Here we focus on public safety nets, which are among the most well-studied mediators of unemployment's effects on wellbeing.

One study found that plant closures led to an almost four-fold greater increase in depression for American workers compared to European workers, who have stronger social safety nets in addition to different cultural values around work \citep{riumallo-herl_job_2014}. This study also found that pre-existing wealth constitutes a significant resilience factor in the US, whereas there was no such interaction effect between wealth and job loss in Europe. This suggests that in settings where there is no comprehensive safety net, adaptation is much more heavily dictated by the financial means of the individual.

By analyzing state-level variations in U.S. unemployment benefits, researchers can more precisely isolate the effect of welfare and retraining policies. This data show that more generous unemployment benefits can significantly reduce the negative effects of unemployment on suicide risk and self-reported health \citep{cylus_generous_2014, cylus_health_2015}.

\subsection{Implications for AI futures}
\label{sec:societal_systemic_implications}

These social factors (in addition to factors of agency and choice) help to explain the notable divergence between the unemployed and retirees, who experience different social expectations around work placed upon them. This dynamic is perhaps most strongly evidenced by the results on ``retiring from unemployment''. The results surveyed in this section have a few important implications for thinking about a world with widespread economic automation.

First, we should not expect one-size-fits-all policies to succeed consistently across societal contexts. The impacts of job loss are heavily moderated by local variables, such as a country's prevailing ``norm to work'' or levels of individualism. For example, in a society with a strong emphasis on the ``norm to work'', a policy like UBI will fail to address the lost social standing. In contrast, UBI might be more sufficient in a society where the ``norm to work'' is weaker. Psychological impacts can also be asymmetric \emph{within} societies, as seen e.g. in the gendered impacts of unemployment.

Second, outcomes are likely to be shaped by whether societal or systemic change is broadly shared or unevenly distributed across populations. That is, they provide a lens for understanding scenarios with differing levels of job displacement. For example, in futures where all of society occupies ``post-labor'' status, expectations around work and the determinants of wellbeing could radically shift (as indicated by the research on ``others' unemployment''), mitigating the psychological penalty of being jobless. Conversely, if AI-driven displacement is unevenly distributed, traditional employment structures will likely persist and continue to shape societal norms. In such a scenario, those who are displaced would likely still face the heavy psychological costs tied to the ``norm to work''. Even those who \emph{retain} their jobs may suffer from decreased wellbeing---as we have seen in cases of high regional unemployment---due to job insecurity, survivor guilt, or a reluctance to leave unsatisfactory roles.

Even in cases where greater levels of displacement could accelerate norm shifts (thus mitigating some psychological effects on individuals), the ``manifest'' or material effects would likely be significant. As reviewed above, social safety nets can play an important role supporting wellbeing in periods of economic change. A key challenge then, is to identify policies that provide the right kind of support, ensuring that material needs are met while also supporting a person's social standing and identity.


\section{Limitations}
\label{sec:limitations}

The analysis provided here is subject to several limitations. The most significant limitation may be that the above studies were carried out in modern largely post-industrial societies---which may differ greatly from the societies of the future. As demonstrated in \S\ref{sec:societal_systemic}, societal norms and systemic circumstances significantly determine the ways in which our wellbeing interacts with our work status, and the value that we place on paid work. In these societies, the ``norm to work'' (\S\ref{sec:social_norms}) is shaped by industrial-era social contracts and work ethic traditions. And the available alternatives to work (\S\ref{sec:alternate_activities}) and social safety nets (\S\ref{sec:safety_nets}) are shaped by specific institutional arrangements.

Nonetheless, we believe that the studies reviewed here have significant value for understanding potential future scenarios, as explained in \S\ref{sec:societal_systemic_implications}. If transitions start to occur in the near or medium-term future, or if the impact of automation is unevenly distributed across society, many people are likely to hold similar beliefs and be impacted in similar ways to the people considered in these studies. Absent fuller societal transformation, the response to AI-initiated change will be impacted by existing norms and structures and based upon similar psychological dynamics to the populations studied above. 

Furthermore, these studies help us identify factors that need to be actively studied and \textit{addressed}, ahead of time, if economic policy is to support human wellbeing both now and in the future. For example, the findings on the ``norm to work'' and on the importance of social safety nets are likely to be highly relevant for shaping future AI-based economic transitions.

Another limitation is that, given the already large scope of this review, we were not able to comprehensively survey all relevant populations. Within modern capitalist societies, we invite greater study of the ``F.I.R.E.'' (Financial Independence, Retire Early) movement. There is also a long tradition in development anthropology that low income and non-western societies can inform us about potential futures; many are societies that already have high structural unemployment and novel experiments with welfare provisions, and would be valuable to study in future work \citep{ferguson_give_2015, Jeske2020-bj, Mains2013-fs, Monteith2021-ad}. We have also not included studies of universal basic income, whose participants do not generally fit our inclusion criteria of ``not working or not needing to work'' \citep{vivalt_employment_2024}, and because these studies are already given high visibility within AI discussions.


\section{Conclusions}
\label{sec:conclusions}

Overall, the empirical evidence demonstrates that wellbeing without traditional paid employment is possible, though by no means guaranteed. By examining modern populations who do not work, we see a clear divergence---involuntary job loss is linked to declines in mental and physical wellbeing; on the other hand, voluntary retirees frequently report high, sustained levels of life satisfaction.

To explain this divergence and to anticipate how human wellbeing might fare under automation, we can return to the three key factors introduced in this review: (1)~individual agency and choice, (2)~alternative sources of work's latent benefits, and (3)~systemic and social contexts.

The first factor highlights that a critical determinant of wellbeing is not just the status of one's employment \textit{per se}, but rather \textit{agency and choice} dictating that status. Retirees typically \textit{choose} to stop working, whereas unemployment is involuntary by definition. This is reflected \textit{within} populations as well---involuntary retirees and involuntary stay-at-home mothers have consistently lower wellbeing than their voluntary counterparts. Lack of agency is important not just in the decision to work, but also in the long run---as demonstrated by financially dependent spouses who may find it difficult to exit unhappy marriages, or Gulf State citizens whose historical reliance on state-provisioned jobs created vulnerabilities when economic conditions shifted. 

The second factor addresses the fact that employment provides a number of ``latent'' benefits beyond financial remuneration, including status, time structure, collective purpose, and cognitive health. Individuals experiencing sudden wealth shocks overwhelmingly choose to remain in the workforce, as demonstrated by data on lottery winners, who instead use their windfalls to purchase flexibility. However, these latent benefits can be found outside of traditional market-driven employment. Wellbeing outcomes from individuals engaged in activities like volunteering, hobbies, or formal care roles---as well as through state-provisioned jobs---demonstrate that alternative sources of purpose and structure can successfully uplift levels of wellbeing.

The third factor emphasizes \textit{social and systemic contexts}, especially social norms and safety nets. For example, the ``norm to work'' significantly moderates the degree to which unemployment is linked to losses in wellbeing. One of the most striking findings in this review regards the phenomenon of ``retiring from unemployment''---when unemployed individuals pass the age of retirement, wellbeing significantly increases, even when there is no change in income or routine. Similarly, the psychological distress of unemployment is buffered when regional unemployment is high---this may herald potential future shifts around the stigma of non-work, in the case of large-scale economic displacement. Institutional support structures also play a vital protective function, with robust welfare and unemployment benefits significantly mitigating the negative wellbeing impacts of unemployment.

In contemporary policy discussions regarding AI-driven job displacement, proposals frequently center on purely material redistribution, most notably ``universal basic income'' (UBI). However, the evidence reviewed here suggests that such interventions fall significantly short. First, they fail to replace the critical ``latent'' benefits of work. Second, they risk creating systemic dependencies that erode human agency, while failing to account for agency and choice during transitions. Third, they neglect the immense psychological pressure of social norms.

Together, these empirical findings point to several major lessons for potential automation-based economic transitions:
\begin{enumerate}
    \item We must prioritize individual agency and choice, both during economic transitions and in the long-term, to avoid dependency scenarios.
    \item A successful transition into an automated economy requires society to furnish alternative sources for the ``latent'' (non-financial) benefits of traditional employment. A portfolio of options (e.g., structured civic roles, volunteering, or state-provisioned job programs) will likely be necessary to accommodate differing individual preferences and needs.
    \item We might expect wellbeing outcomes to be shaped by profound cultural shifts in the longer term regarding human value and purpose. Nonetheless, in the near term, we will likely see the persistence of entrenched norms surrounding the moral imperative to work.
    \item Navigating such transitions will therefore require comprehensive institutional support and robust safety nets, to ensure that those affected by technological automation can maintain both their material livelihoods and their psychological wellbeing.
\end{enumerate}

\vspace{.2em}
\section*{Acknowledgments}
We are extremely grateful to Claudia Strauss, Murray Shanahan, Haydn Belfield, Shamil Chandaria, and Julian Jacobs for helpful feedback and discussions that shaped this paper.

\bibliographystyle{plainnat}
\bibliography{references}

\appendix

\begin{table*}
\section*{Appendix}
\vspace{10pt}
\renewcommand{\arraystretch}{1.5}
\begin{tabular}{>{\raggedright\arraybackslash}p{0.15\textwidth} p{0.2\textwidth} p{0.28\textwidth} p{0.22\textwidth}}
\toprule
\textbf{} & \textbf{Description} & \textbf{Advantages \& Disadvantages} & \textbf{Examples} \\
\hline

\textbf{Cross-sectional studies} & Observations drawn from a single point in time. Individuals can be drawn from different populations (e.g. different countries or employment status). & Allows researchers to capture a snapshot of current conditions or to compare across different populations. However, they cannot establish causal direction or temporal precedence, and are vulnerable to unobserved individual differences and confounding variables. Generally the weakest design of those listed here. & Comparing different employed vs unemployed individuals at a single point in time \citep{gallie_unemployment_1999}. \\

\textbf{Longitudinal (panel) datasets} & 
Track the same individuals over a period of time. & 
Provides stronger support for causal inferences through temporal precedence, and controls for individual differences. However, datasets are significantly more challenging to obtain. & 
Comparing the behavior of the same individuals when they are employed vs unemployed \citep{rozer_does_2020}. \\

\textbf{Randomized controlled trials (RCTs) }& Participants are randomly assigned to receive a particular treatment (or to a control group). & Considered the gold standard for causal inference, by reducing selection bias and confounding factors. However, they can be difficult, expensive, or sometimes unethical to implement. & Randomly assigning subjects to participate in volunteering \citep{jongenelis_longitudinal_2022} or in a youth job program \citep{aizer_youth_2020}. \\

\textbf{Natural \& quasi-experiments} & Leverage naturally occurring events or variations to study exogenous shocks. & Strong causal inference; isolates the impacts of an event from individual selection factors (e.g., preexisting physical or mental health, which may affect an individual's tendency towards job loss). & Lottery wins (random wealth shocks); the closure of an entire firm (involuntary job loss); heterogeneity of retirement policies across countries \citep{rohwedder_mental_2010}. \\

\textbf{Instrumental variable \mbox{analysis} }& A statistical technique using an external variable (instrument) to estimate causal relationships. & A powerful statistical technique for estimating causal relationships when randomized trials are not possible. However, findings can only be interpreted as applying to the specific individuals whose choices were changed by the instrument. & Using eligibility for Social Security as an instrument to evaluate the effect of retirement \citep{yemiscigil_effects_2021}. \\

\textbf{Meta-analyses} & Studies that systematically combine and analyze quantitative data from multiple independent studies. & Greatly improves statistical power for identifying trends. Resolves uncertainties or mixed results. & Consolidating and aggregating data from dozens of longitudinal studies to quantify the impact of unemployment on wellbeing \citep{gedikli_relationship_2023} \\
\bottomrule

\end{tabular}
\caption{Types of studies and analytical techniques}
\label{table:studies_and_techniques}
\end{table*}

\end{document}